\RequirePackage{fix-cm}
\documentclass[smallextended]{svjour3}       % onecolumn (second format)
\smartqed  % flush right qed marks, e.g. at end of proof
\usepackage{graphicx}
\usepackage{amsmath}
\usepackage{subfigure}
\usepackage{braket}
\usepackage{pgfplots}
\usepackage{csquotes}
\usepackage{float}
\usepackage{multirow}
\usepackage{dcolumn}
\usepackage[colorlinks=true, citecolor=blue]{hyperref}
\usepackage{hhline}
\usepackage{amssymb}
\begin{document}

\title{Experimental realization of quantum teleportation using coined quantum walks} 
\author{Yagnik Chatterjee \and Vipin Devrari \and Bikash K. Behera \and Prasanta K. Panigrahi}

\institute{Yagnik Chatterjee \at Department of Physics and Astronomy, National Institute of Technology Rourkela, Rourkela 769008, Odisha, India\\ \email{418ph2107@nitrkl.ac.in} \and Vipin Devrari \at Department of Physics and Astronomy, National Institute of Technology Rourkela, Rourkela 769008, Odisha, India \\ \email{418ph2109@nitrkl.ac.in} \and Bikash K. Behera \at  Department of Physical Sciences, Indian Institute of Science Education and Research Kolkata, Mohanpur 741246, West Bengal, India \\ \email{bkb18rs025@iiserkol.ac.in} \and Prasanta K. Panigrahi \at Department of Physical Sciences, Indian Institute of Science Education and Research Kolkata, Mohanpur 741246, West Bengal, India \\ \email{pprasanta@iiserkol.ac.in}}

\date{Received: date / Accepted: date}

\maketitle

\begin{abstract}
The goal of teleportation is to transfer the state of one particle to another particle. In coined quantum walks, conditional shift operators can introduce entanglement between position space and coin space. This entanglement resource can be used as a quantum channel for teleportation, as proposed by Wang, Shang and Xue [Quantum Inf. Process. \textbf{16}, 221 (2017)]. Here, we demonstrate the implementation of quantum teleportation using quantum walks on a five-qubit quantum computer and a 32-qubit simulator provided by IBM quantum experience beta platform. We show the teleportation of single-qubit, two-qubit and three-qubit quantum states with circuit implementation on the quantum devices. The teleportation of Bell, W and GHZ states has also been demonstrated as special cases of the above states. 
\end{abstract}


\section{Introduction}
 
The quantum teleportation \cite{qtc_tele1,qtc_tele2,qtc_tele3,qtc_tele4} scheme was originally proposed by Bennett \emph{et al.} \cite{qtc_bennett}, and it has now become a key technology in many quantum information processing tasks. In general, there are two crucial processes in teleportation protocol, namely preparation of entanglement states and joint measurement. In some sense, the goal of teleportation is to transfer the state of one particle to another particle. 

Quantum walks are quantum counterparts of classical random walks. Classical random walks have been successfully adopted to develop classical algorithms. This has resulted in a huge interest in understanding the properties of quantum walks to create quantum algorithms which are faster than classical algorithms. Quantum walks have extensive applications in the field of computer science. In addition, the study of quantum walks is relevant to building methods to test the `quantumness' of emerging technologies for the creation of quantum computers as well as for quantum simulation of natural phenomena. In coined quantum walks \cite{qtc_qwalk1,qtc_qwalk2}, conditional shift operators can introduce entanglement between position space and coin space. Wang, Shang and Xue \cite{qtc_wang} have proposed a scheme to use this entanglement resource as a quantum channel for teleportation. Perhaps the most intriguing aspect of teleportation by quantum walks is that maximal entanglement \cite{qtc_ent1,qtc_ent2,qtc_ent3} resource is not necessarily prepared beforehand. It can be generated naturally by two conditional shift operators during the process of the walk. 

IBM Q \cite{qtc_ibmq} is a superconducting qubit based operating system which offers open global access to a wide class of researchers and has found significant applications \cite{qtc_behera1,qtc_behera2,qtc_behera3,qtc_behera4,qtc_behera5,qtc_behera6} in an user-friendly interface. In this paper, we use a five-qubit quantum computer as well as a 32-qubit simulator provided by IBM to implement teleportation using quantum walks on graphs. Transfer procedure is implemented between two coin spaces only by two-step quantum walk. Any unknown n-qubit state can be teleported by quantum walks on a cycle with n-vertices and n-directed edges at each vertex. Here, we prepare quantum circuits and experimentally show the teleportation of any general single-qubit, two-qubit and three-qubit state. As special cases, we have shown the teleportation of Bell \cite{qtc_bell1,qtc_bell2}, GHZ \cite{qtc_ghz1,qtc_ghz2,qtc_ghz3,qtc_ghz4} and W \cite{qtc_w1,qtc_w2} states.

The paper is organized as follows. Section \ref{qtc_Sec2} introduces coined quantum walks on graphs. Section \ref{qtc_Sec3} explains the general scheme for teleportation that we have used. Section \ref{qtc_Sec4}, Section \ref{qtc_Sec5} and Section \ref{qtc_Sec6} cover the theory and implementation of teleportation of single-qubit, two-qubit and three-qubit states using quantum walks in detail. Finally, we conclude in Section \ref{qtc_Sec7}. 

\section{Coined Quantum Walks on Graphs \label{qtc_Sec2}}
  
Quantum walks are the quantum analogue of classical random walks. We consider the definition of coined quantum walks on graphs. Let $G(V,E)$ be a graph, and $H_v$ be the Hilbert space spanned by states $\ket{v}$, where vertex $v \in V$ . At each vertex $j$, there are several directed edges with labels pointing to the other vertices. The coin space $H_c$ is spanned by states $\ket{a}$, where $a \in \{0,. . , d-1\}$. They are the labels of the directed edges. The conditional shift operation between $H_v$ and $H_c$ is $T = \sum_{jk} \ket{k}\bra{j} \otimes \ket{a}\bra{a}$, where the label a directs the edge $j$ to $k$. This quantum walk has a single coin space. However, it is quite possible to have quantum walks with more than one coin, as shall be demonstrated in our teleportation scheme. 

\section{Scheme for Teleportation \label{qtc_Sec3}} 

The general scheme used here is quite straightforward. We have a sender and a receiver, say Alice and Bob. Alice and Bob have a coin each. Alice's coin is in the unknown state $\ket{\phi}$ which she needs to send to Bob. The quantum walk is performed on a graph having d-vertices with each vertex having d-directed edges, where d is the dimension of the coin space. 

\begin{figure}[H]
\centering
\includegraphics[scale=0.2]{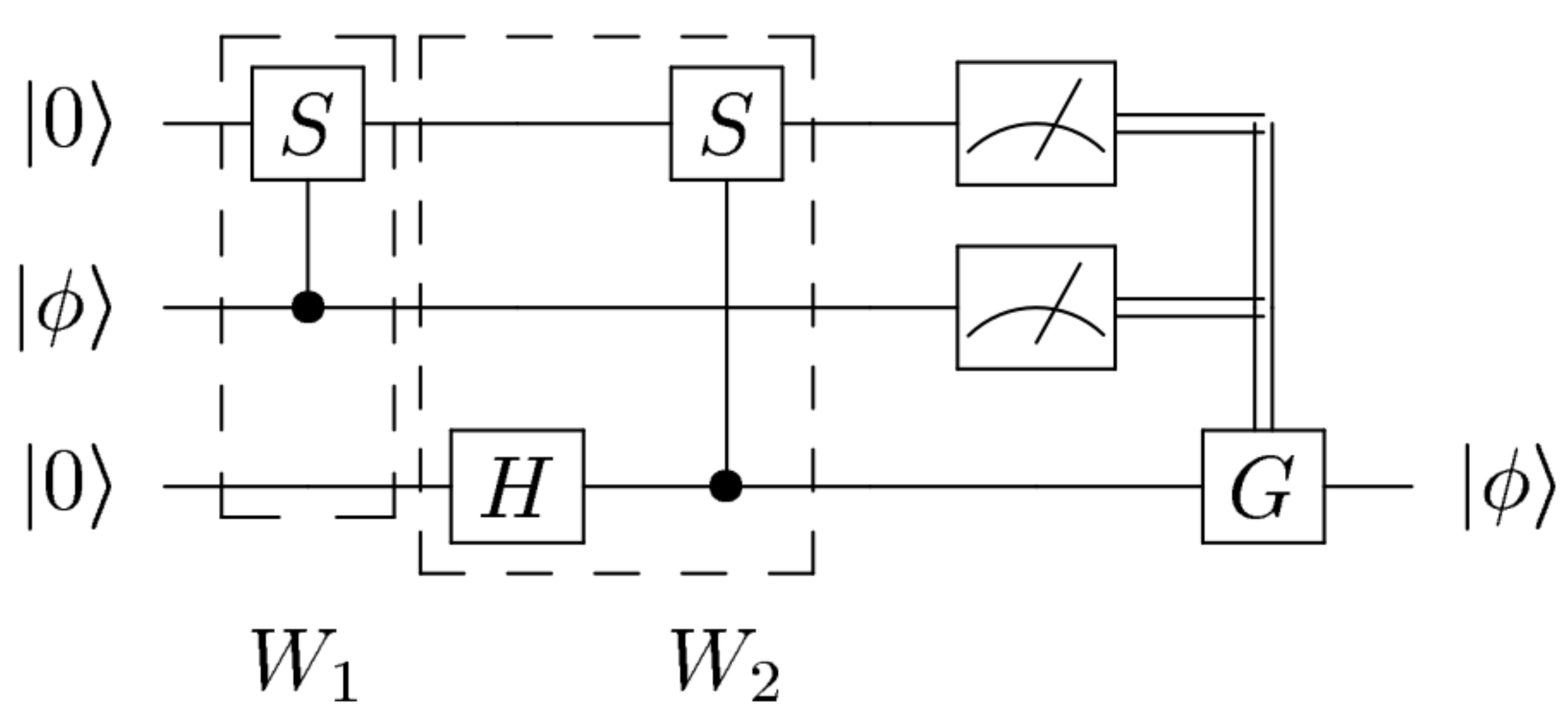}
\caption{\textbf{Teleportation protocol}. \emph{General scheme for teleportation of any d-dimensional state $\ket{\phi}$ (qudit). Position space is represented by the top wire. The centre and bottom wires represent Alice's coin and Bob's coin respectively. Here, $S$ represents the shift operation.}}
\label{qtc_Fig1}
\end{figure}

In the first step, Alice applies the conditional shift operation on the position space using her coin. This step entangles the position space and Alice's coin space. In the second step, Bob will first apply a Hadamard gate on his coin and then apply the conditional shift operator on the position space using his coin. It should be noted that in case of a qudit with $d>2$, the Hadamard will effectively split up every basis state into all possible states with equal probability, although the phase of each state might differ. These two steps ensure that the position space and both coin spaces are entangled with each other. Alice now performs a measurement on her qudit and the position qudit in an appropriate basis. She sends the classical information to Bob, who can then recover the state $\ket{\phi}$ by applying appropriate operations $G$.

\section{Single Qubit Teleportation} \label{qtc_Sec4}

\subsection{Theoretical Description}

\begin{figure}[H]
\centering
\includegraphics[scale=0.2]{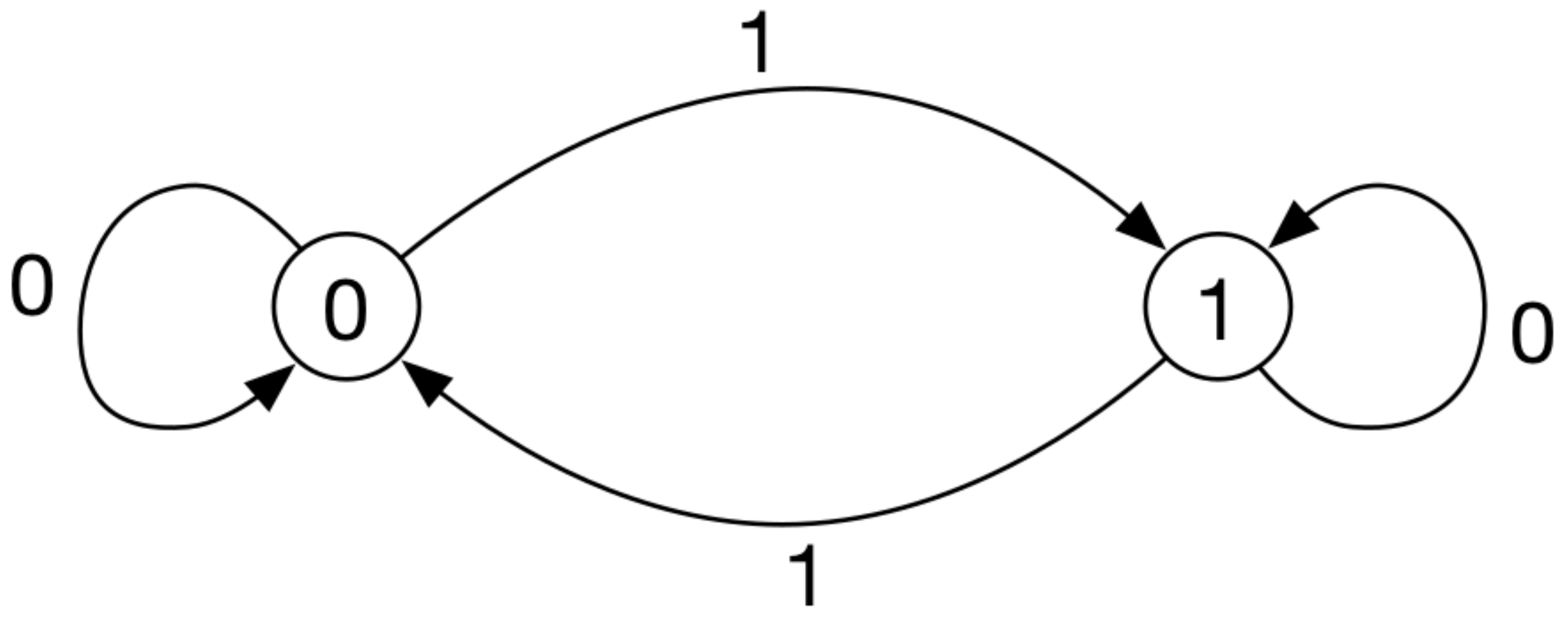}
\caption{\textbf{Complete graph with two vertices.} \emph{The vertices represent the states $\ket{0}$ and $\ket{1}$. If the control qubit is 0, then the directed edge points to the same state and if it is 1, the directed edge points to the other state.}}
\label{qtc_Fig2}
\end{figure}

For single-qubit teleportation, the quantum walk is carried out on a graph with two vertices. At each vertex, there are two directed edges as can be seen in Fig. \ref{qtc_Fig2}. The dimension of the coin spaces and the position space is obviously $d=2$, since we need to teleport a single-qubit state. The position remains in the same state if the control qubit is $\ket{0}$ and the state is flipped otherwise. Let the initial state be (Eq. \eqref{qtc_Eq1}),
\begin{equation}
    \ket{\psi_0}=\ket{0} \otimes \ket{\phi} \otimes \ket{0}
    \label{qtc_Eq1}
\end{equation}
where, $\ket{\phi}= a\ket{0}+b\ket{1}$, and $|a|^2+|b|^2=1$. In the first step (Fig. \ref{qtc_Fig1}), a controlled-S$_{21}$ operation is applied from the second qubit to the first qubit, which results (Eq. \eqref{qtc_Eq2}), 
\begin{equation}
    \ket{\psi_1} =  a\ket{000}+ b\ket{110}
    \label{qtc_Eq2}
\end{equation}

It is to be noted that the shift operator, S used here is a NOT operation, and controlled-S$_{ij}$ denotes the application of S operation on the $j^{th}$ qubit where $i^{th}$ qubit acts as the control qubit. In the second step, Bob will first apply Hadamard to the third qubit and a controlled-S$_{31}$ operation, resulting (Eq. \eqref{qtc_Eq3})

\begin{equation}
\ket{\psi_2} = \ket{0} \otimes \frac{a\ket{00} + b\ket{11}}{\sqrt{2}} + \ket{1} \otimes \frac{a\ket{01} + b\ket{10}}{\sqrt{2}}
\label{qtc_Eq3}
\end{equation}

Alice now performs measurement on the position qubit in $\ket{0}$ and $\ket{1}$ basis and on her own qubit (coin) in $\ket{+}$ and $\ket{-}$ basis. Let us call these measurements $M_1$ and $M_2$ respectively. She then sends these measured classical bits to Bob. Depending on what he receives, he can perform the following operations to recover the state $\ket{\phi}$.

\begin{table}[H]
\caption{\emph{Measurement results and corresponding revise operations}}
\centering
\begin{tabular}{|c|c|c|}
 \hline
 $M_1$ & $M_2$
 & Bob's revise operation\\
 \hline
0 & 0 & I\\
0 & 1 & Z\\
1 & 0 & X\\
1 & 1 & ZX\\
 \hline
\end{tabular}
\label{qtc_Tab1}
\end{table}

\subsection{Implementation and Results} 

\begin{figure}[H]
    \centering
    \includegraphics[scale=0.2]{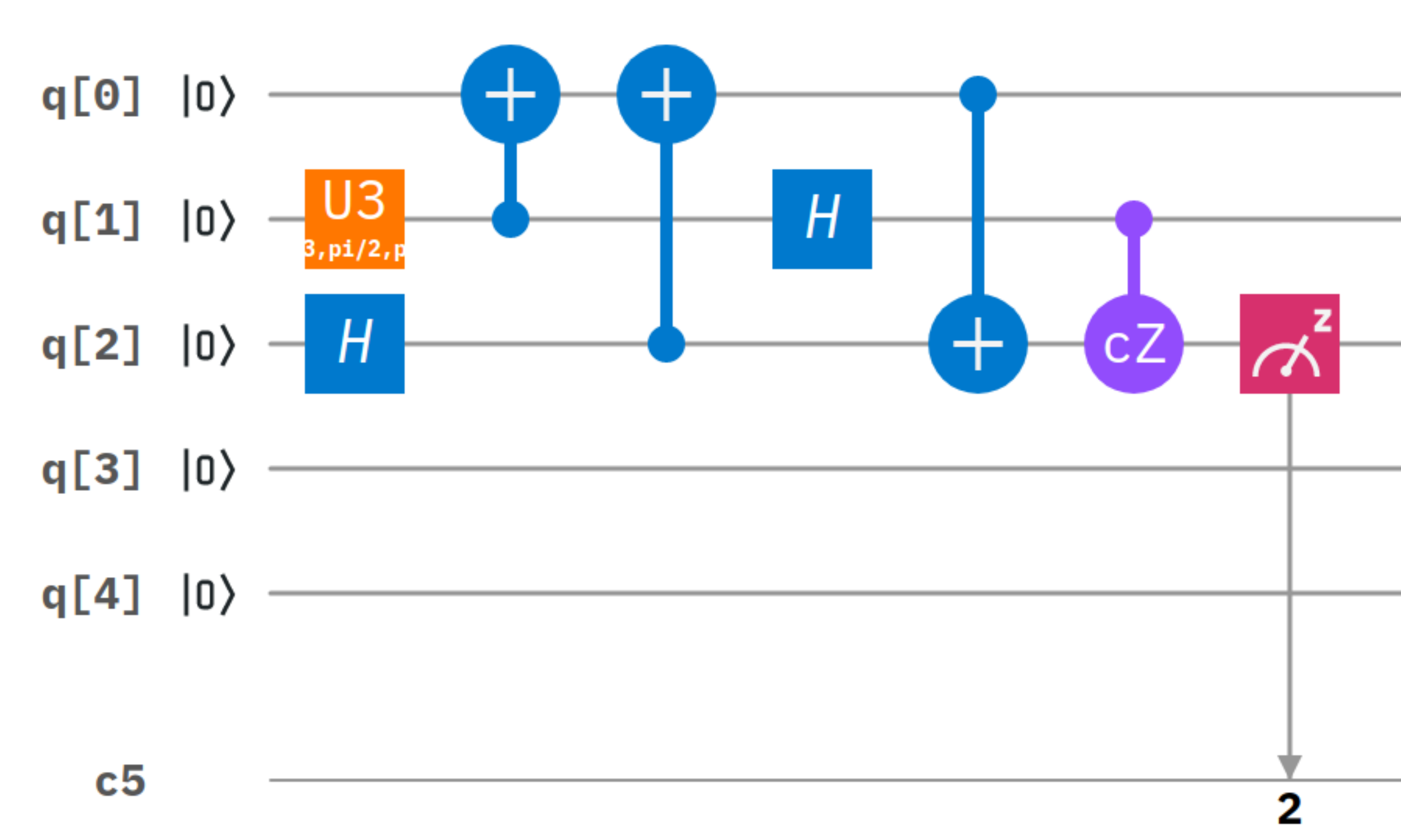}
    \caption{\textbf{Quantum circuit for the teleportation of single qubit state.} \emph{The qubit q[0] represents the position space. The qubit q[1] represents Alice's coin and q[2] represents Bob's coin.} }
     \label{qtc_Fig3}
\end{figure}

\begin{figure}[H]
    \centering
    \includegraphics[scale=0.52]{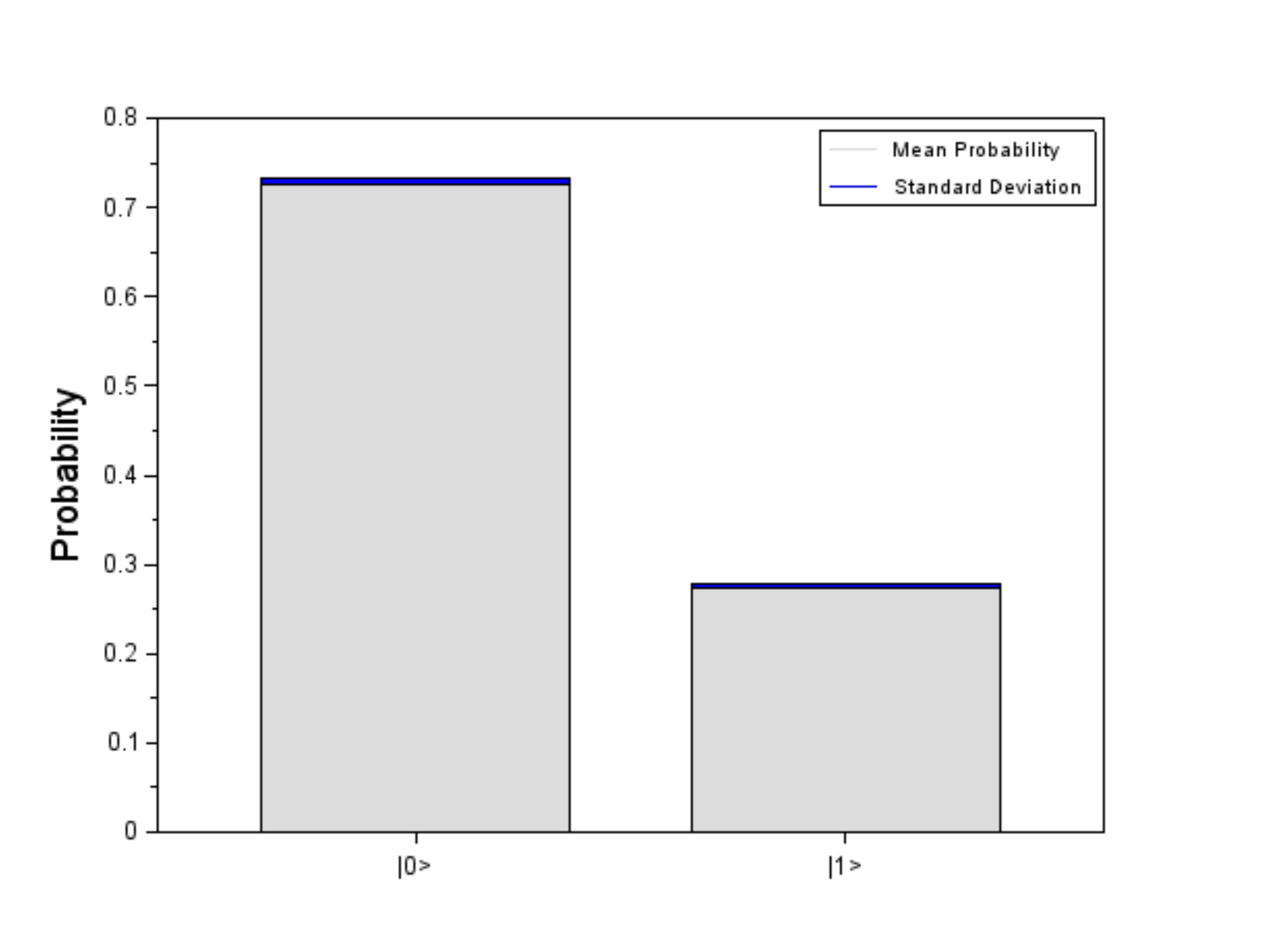}
    \caption{\textbf{Histogram.} \emph{This histogram shows the mean probabilities for $\ket{0}$ and $\ket{1}$ states and their standard deviations.} }
     \label{qtc_Fig4}
\end{figure}

The initial state, prepared by using the $U3(\pi/3,\pi/2,\pi/2)$ gate is $\ket{\phi}=\frac{\sqrt{3}}{2}\ket{0} + \frac{1}{2}\ket{1}$. The five-qubit IBM quantum computer \emph{ibmqx2} has been used for this experiment. Table \ref{qtc_Tab2} shows the data taken for 10 runs, each run having 8192 shots. Fig. \ref{qtc_Fig4} shows the graphical representation of the same. The measurements are in $Z$-basis.

\begin{table}[H]
\caption{\emph{The table shows the results of teleportation on a five-qubit quantum computer (ibmqx2). Number of shots used for each run is 8192.}}
\centering
\begin{tabular}{|c|c|c|}
\hline
Runs & Probability of $\ket{0}$ & Probability of $|1\rangle$\\
\hline
Run 1 & 0.726 & 0.274\\
Run 2 & 0.738 & 0.262\\
Run 3 & 0.730 & 0.270\\
Run 4 & 0.720 & 0.280\\
Run 5 & 0.725 & 0.275\\
Run 6 & 0.724 & 0.276\\
Run 7 & 0.720 & 0.280\\
Run 8 & 0.725 & 0.275\\
Run 9 & 0.729 & 0.271\\
Run 10 & 0.727 & 0.273\\
\hline
\end{tabular}
\label{qtc_Tab2}
\end{table}

\subsection{Quantum State Tomography}

We now proceed to carry out state tomography \cite{qtc_tomo1} to check how well the quantum states are teleported in our experiment. In this process, by comparing both the theoretical and experimental density matrices of a quantum state, the accuracy of implementation can be tested. The theoretical and experimental density matrices of the given state are given by the following equations (Eqs. \eqref{qtc_Eq4} and \eqref{qtc_Eq5}):

\begin{equation}
    \rho^{T} = \ket{\psi}\bra{\psi}
\label{qtc_Eq4}
\end{equation}

\begin{equation}
     \rho^{E}=\frac{1}{2} (I+\langle X \rangle X + \langle Y \rangle Y + \langle Z \rangle Z)
     \label{qtc_Eq5}
\end{equation}

From our initial state $\ket{\psi}=\frac{\sqrt{3}}{2}\ket{0} + \frac{1}{2}\ket{1}$, and the data in Table \ref{qtc_Tab2}, theoretical ($\rho^T$) and experimental ($\rho^E$) density matrices are calculated (Eqs. \eqref{qtc_Eq6} and \eqref{qtc_Eq7}), as given below.

\begin{equation}
 \rho^{T} =
\begin{pmatrix}
      0.75 & 0.433\\
     0.433 & 0.25
\end{pmatrix}
\label{qtc_Eq6}
\end{equation}

\begin{equation}
\rho^{E} =
\begin{pmatrix}
      0.7264 & 0.2912\\
     0.2912 & 0.2736
\end{pmatrix}
+
\imath
\begin{pmatrix}
      0 & -0.0024\\
     0.0024 & 0
\end{pmatrix}
\label{qtc_Eq7}
\end{equation}

\begin{figure}[H]
\centering
\includegraphics[scale=0.21]{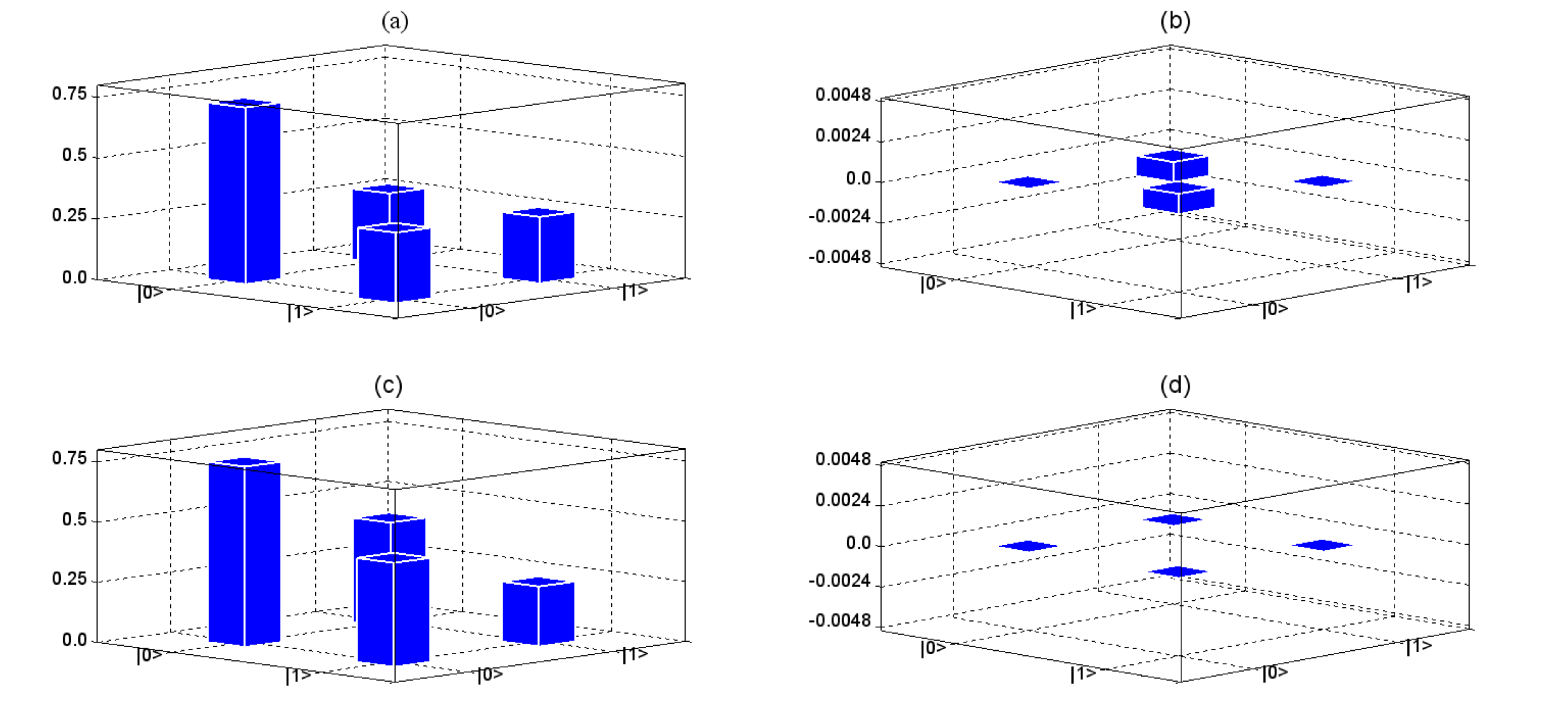}
\caption{\textbf{Real and imaginary parts of theoretical and experimental density matrices for single-qubit quantum teleportation by coins.} \emph{ (a): Real part of experimental density matrix, (b): Imaginary part of experimental density matrix; (c): Real part of theoretical density matrix, (d): Imaginary part of theoretical density matrix.}}
\label{qtc_Fig5}
\end{figure}

Fidelity measures the overlap between two density matrices and hence quantifies the closeness of theoretical and experimental quantum states obtained as output. It can be calculated using the following formula (Eq. \eqref{qtc_Eq8}):

\begin{equation}\label{qtc_Eq8}
F(\rho^{T},\rho^{E})= \Bigg[Tr\bigg(\sqrt{\sqrt{\rho^{T}}\rho^{E}\sqrt{\rho^{T}}}\bigg)\Bigg]^2
\end{equation}

From Fig. \ref{qtc_Fig4}, the accuracy can be easily checked by comparing the theoretical and experimental density matrices. Fidelity of the experimental result is found to be $F= 0.865$. 

\section{Two-Qubit Teleportation} \label{qtc_Sec5}

\subsection{Theoretical Description}\label{qtc_Sec5.1}

For two-qubit teleportation, the quantum walk is carried out on a complete graph with four vertices. At each vertex, there are four directed edges. Let the vertices represent the states $\ket{00}$, $\ket{01}$, $\ket{10}$ and $\ket{11}$ in position space. Since Alice needs to teleport a two-qubit state, both Alice and Bob need coins with two qubits. Any arbitrary two-qubit state can be written as, $$\ket{\phi}= a\ket{00}+b\ket{01}+c\ket{10}+d\ket{11}$$. 
   
Let the initial state be (Eq. \eqref{qtc_Eq9}) 
\begin{equation}\label{qtc_Eq9}
    \ket{\psi_0}=\ket{00} \otimes \ket{\phi} \otimes \ket{00}
\end{equation}
where, $\ket{\phi}= \sum_{i,j\in \{0,1\}} a_{ij}\ket{i\,j}$. Therefore the initial state can be written as,

\begin{equation}\label{qtc_Eq10}
    \ket{\psi_0} = \ket{00} \otimes  \sum_{i,j\in \{0,1\}} a_{ij}\ket{i\,j} \otimes \ket{00}
\end{equation}

For simplification, we use the following notation: $\ket{i\,j}= \ket{2i+j}$ where $ i,j \in \{0,1\}$. Hence the basis states $\ket{00}$, $\ket{01}$, $\ket{10}$ and $\ket{11}$ will be represented by $\ket{0}$, $\ket{1}$, $\ket{2}$ and $\ket{3}$ respectively. This is equivalent to a $d=4$ qudit system. In this representation, the initial state can be written as,

\begin{equation}\label{qtc_Eq11}
    \ket{\psi_0} = \ket{0} \otimes  \sum_{k=0}^{3} a_{k}\ket{k} \otimes \ket{0}
\end{equation}

The action of the shift operator can be defined as follows. If the state of the qudit (Alice's or Bob's coin as applicable) is $\ket{k}$ and the state of the position space is $\ket{l}$ where $k,l\in\{0,1,2,3\}$, then the shift operator $T$ is,

\begin{equation}\label{qtc_Eq12}
    T_{lk}= \ket{(l+k)\, mod\, 4}\bra{l} \otimes \ket{k}\bra{k}
\end{equation}

In the first step, Alice applies the conditional shift operator on the position space using her coin. After the first step (Fig. \ref{qtc_Fig1}), the state becomes 
\begin{equation}\label{qtc_Eq13}
    \ket{\psi_1} =  \sum_{k=0}^{3} a_k\ket{k\,k\,0}
\end{equation}

In the second step, Bob will first apply Hadamard to the third qubit, making the state
\begin{equation}\label{qtc_Eq14}
    \ket{\psi_2} = \frac{1}{2} \sum_{k=0}^{3} \sum_{m=0}^{3}a_k\ket{k\,k\,m}
\end{equation}
and then apply the shift operator using his coin, that results
\begin{equation}\label{qtc_Eq15}
    \ket{\psi_3} =  \frac{1}{2}\sum_{k=0}^{3} \sum_{m=0}^{3}a_k\ket{(k+m)\,mod\,4} \otimes \ket{k\,m}
\end{equation}

Alice now performs measurement on the position qudit in $\ket{0}$, $\ket{1}$, $\ket{2}$ and $\ket{3}$ basis. If upon measurement she gets the state $\ket{q}$, $q \in \{0,1,2,3\}$, then the state collapses to,
\begin{equation}\label{qtc_Eq16}
    \ket{\psi_4} =  \frac{1}{2}\sum_{k=0}^{3}a_k\ket{k} \otimes \ket{(q-k)\,mod\,4}
\end{equation}
Alice has to measure the state of her coin in $\ket{++}$, $\ket{+-}$, $\ket{-+}$ and $\ket{--}$ basis states. Thus Alice applies the Hadamard gate on her qubits and the state becomes,
\begin{equation}\label{qtc_Eq17}
    \ket{\psi_5} =  \frac{1}{4}\sum_{k=0}^{3}\sum_{i=0}^{3}a_k (-1)^{i.k}\ket{i} \otimes \ket{(q-k)\,mod\,4}
\end{equation}
where $i.k$ is the bitwise \emph{and} operation between the binary representations of $i$ and $k$. Upon measurement of her coin, if Alice gets $\ket{p}$, the state now becomes 
\begin{equation}\label{qtc_Eq18}
    \ket{\psi_6} =  \frac{1}{4}\sum_{k=0}^{3}a_k (-1)^{p.k} \ket{(q-k)\,mod\,4}
\end{equation}

This is the state that Bob receives when Alice sends her the classical bits $q$ and $p$. To finally recover the original state, Bob has to perform the following operation on $\ket{\psi_6}$.

\begin{equation}\label{qtc_Eq19}
    U_{pq} = \sum_{j=0}^{3} (-1)^{p.j} \ket{j}\bra{(q-j)\,mod\,4}
    \end{equation}
\begin{eqnarray}
    U_{pq} \ket{\psi_6} &=& \frac{1}{4}\sum_{k=0}^{3} \sum_{j=0}^{3}a_k (-1)^{p.k} (-1)^{p.j} \ket{j}\braket{(q-j)\,mod\,4|(q-k)\,mod\,4} \nonumber \\
    &=& \frac{1}{4}\sum_{k=0}^{3} \sum_{j=0}^{3}a_k (-1)^{p.k} (-1)^{p.j} \ket{j} \delta{jk} \nonumber\\
     &=& \frac{1}{4}\sum_{j=0}^{3}a_j (-1)^{p.j} (-1)^{p.j} \ket{j} \nonumber\\
     &=& \frac{1}{4}\sum_{j=0}^{3}a_j \ket{j}\\
\end{eqnarray}

which is the required original state.

\subsection{Implementation and Results}

\begin{figure}[H]
    \centering
    \includegraphics[scale=0.3]{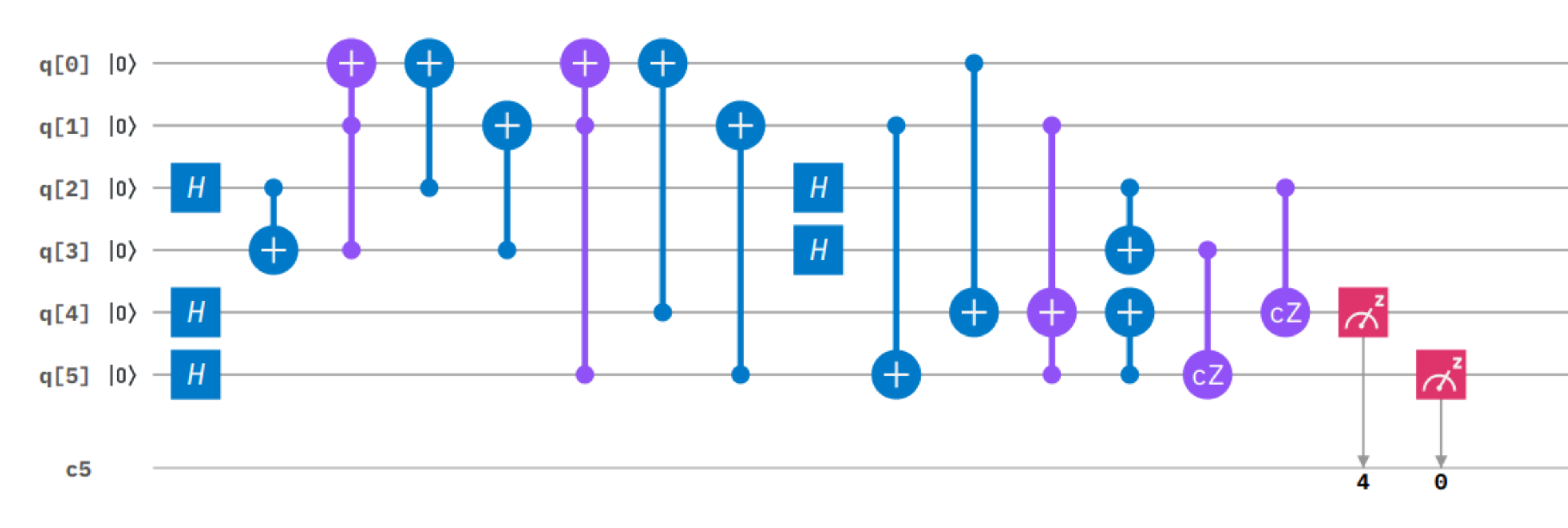}
    \caption{\textbf{Experimental circuit to teleport a Bell State.} \emph{The qubits q[0] and q[1] represent the position space. The qubits q[2] and q[3] represent Alice's coin and q[4] and q[5] represent Bob's coin.} }
     \label{qtc_Fig6}
\end{figure}

\begin{figure}[H]
    \centering
    \includegraphics[scale=0.52]{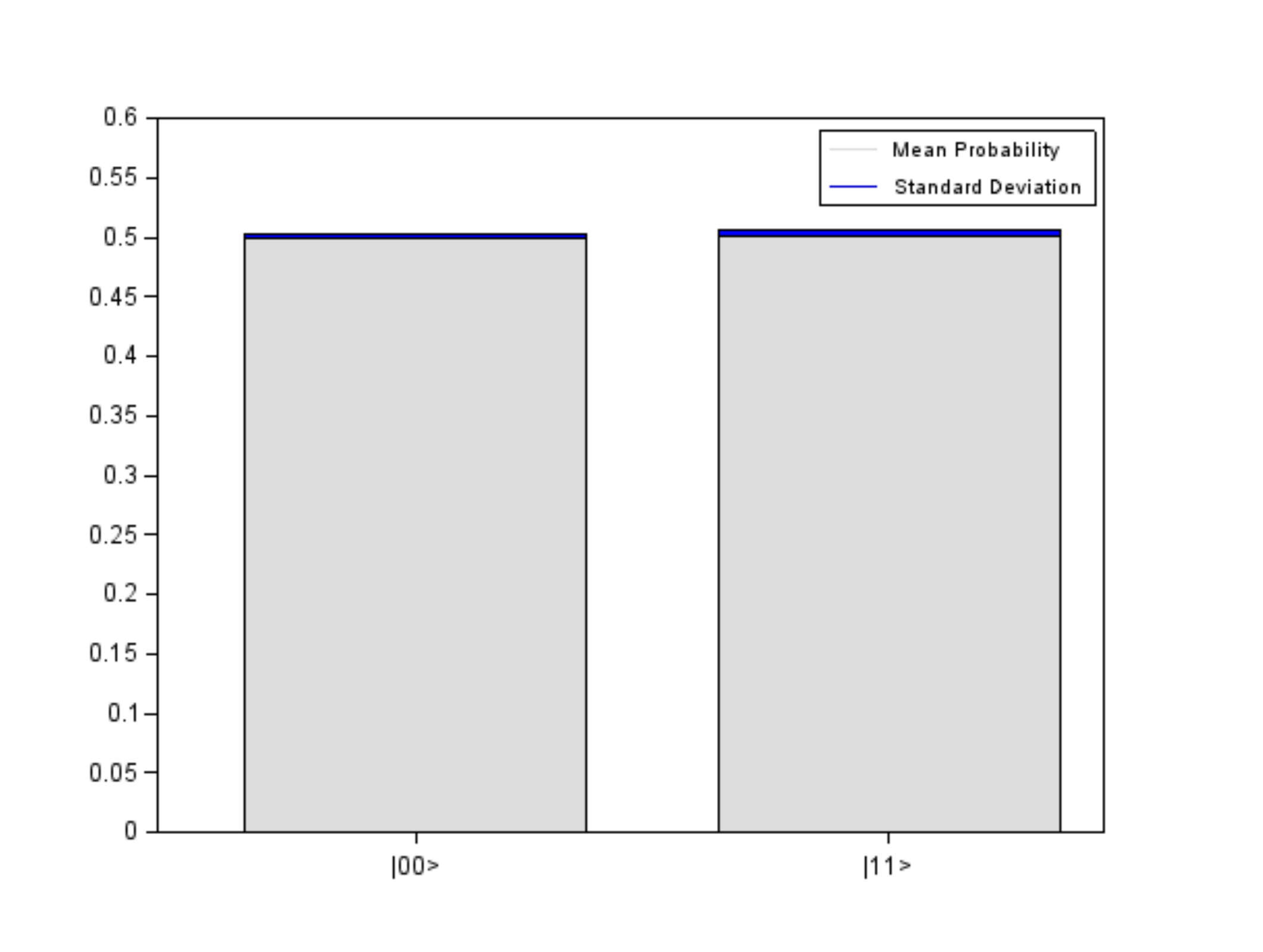}
    \caption{\textbf{Histogram.} \emph{The histogram shows the mean probabilities for $\ket{00}$ and $\ket{11}$ states and their standard deviations.} }
     \label{qtc_Fig7}
\end{figure}

 The above circuit teleports a Bell State as an example of a two-qubit state. The circuit is implemented on a 32-qubit simulator provided by IBM. For each simulation, 8192 shots have been used. Fig. \ref{qtc_Fig7} shows the graphical representation of data taken for 10 simulations. The measurements are in $Z$-basis.

\section{Three-Qubit Teleportation}\label{qtc_Sec6}

\subsection{Theoretical Description}

For three-qubit teleportation, the quantum walk is carried out on a complete graph with eight vertices. At each vertex, there are eight directed edges. Let the vertices represent the states $\ket{000}$, $\ket{001}$, $\ket{010}$, $\ket{011}$, $\ket{100}$, $\ket{101}$, $\ket{110}$ and $\ket{111}$ in position space. Since Alice needs to teleport a three-qubit state, both Alice and Bob need coins with three qubits. Any arbitrary three-qubit state can be written as, $\ket{\phi}= a\ket{000}+b\ket{001}+c\ket{010}+d\ket{011}+e\ket{100}+f\ket{101}+g\ket{110}+h\ket{111}$. 
   
Let the initial state be 
\begin{equation}\label{qtc_Eq24}
    \ket{\psi_0}=\ket{000} \otimes \ket{\phi} \otimes \ket{000}
\end{equation}
where, $\ket{\phi}= \sum_{i,j,k\in \{0,1\}} a_{ijk}\ket{i\,j\,k}$

Therefore the initial state can written as,
\begin{equation}\label{qtc_Eq26}
    \ket{\psi_0} = \ket{000} \otimes  \sum_{i,j,k\in \{0,1\}} a_{ijk}\ket{i\,j\,k} \otimes \ket{000}
\end{equation}

For simplification, we use the following notation: $\ket{i\,j\,k}= \ket{4i+2j+k}$ where $ i,j,k \in \{0,1\}$. This makes the state equivalent to a $d=8$ qudit.

In this representation, the initial state can be written as,
\begin{equation}\label{qtc_Eq27}
    \ket{\psi_0} = \ket{0} \otimes  \sum_{k=0}^{7} a_{k}\ket{k} \otimes \ket{0}
\end{equation}

The form of Eq. \eqref{qtc_Eq27} is almost exactly the same as that of Eq. \eqref{qtc_Eq11}, the only difference being that the index $k$ goes up to $7$ instead $3$. Therefore, by progressing in the same way as section \ref{qtc_Sec5.1}, we can teleport any arbitrary three-qubit state. 

\subsection{Implementation and Results}

\begin{figure}[]
\begin{subfigure}
    \centering
    \includegraphics[scale=0.5]{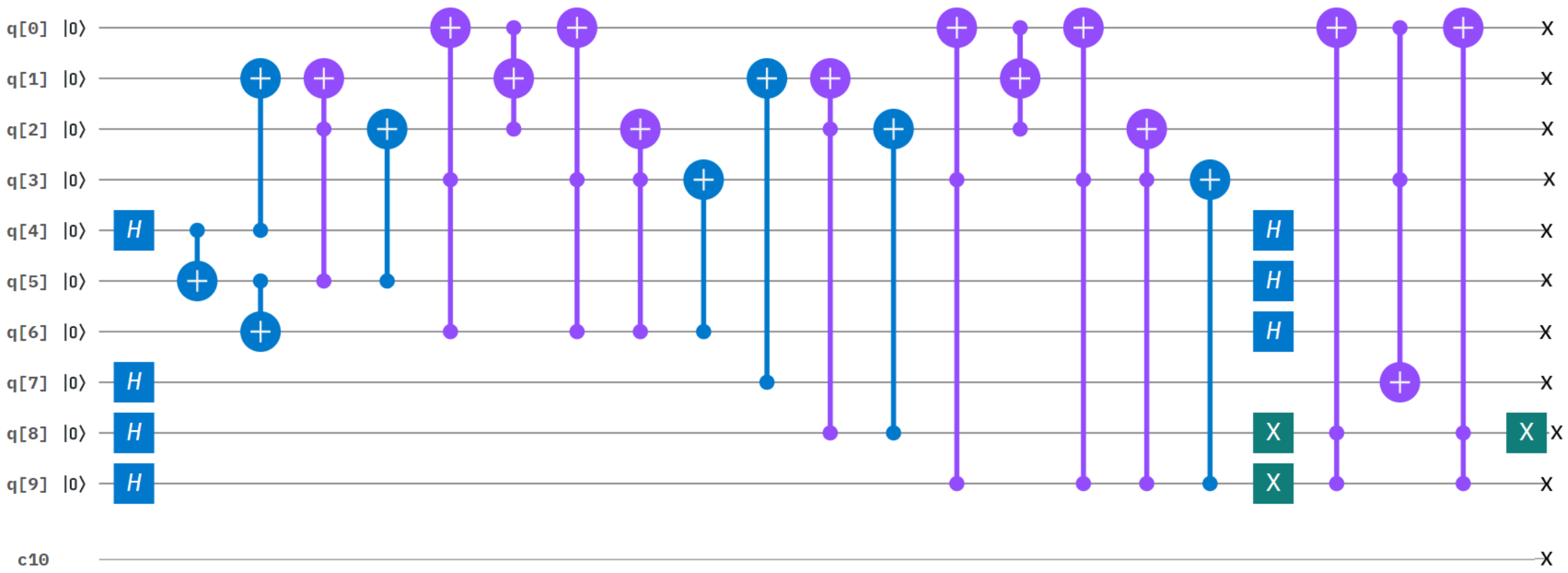}
\end{subfigure}
\begin{subfigure}
    \centering
    \includegraphics[scale=0.5]{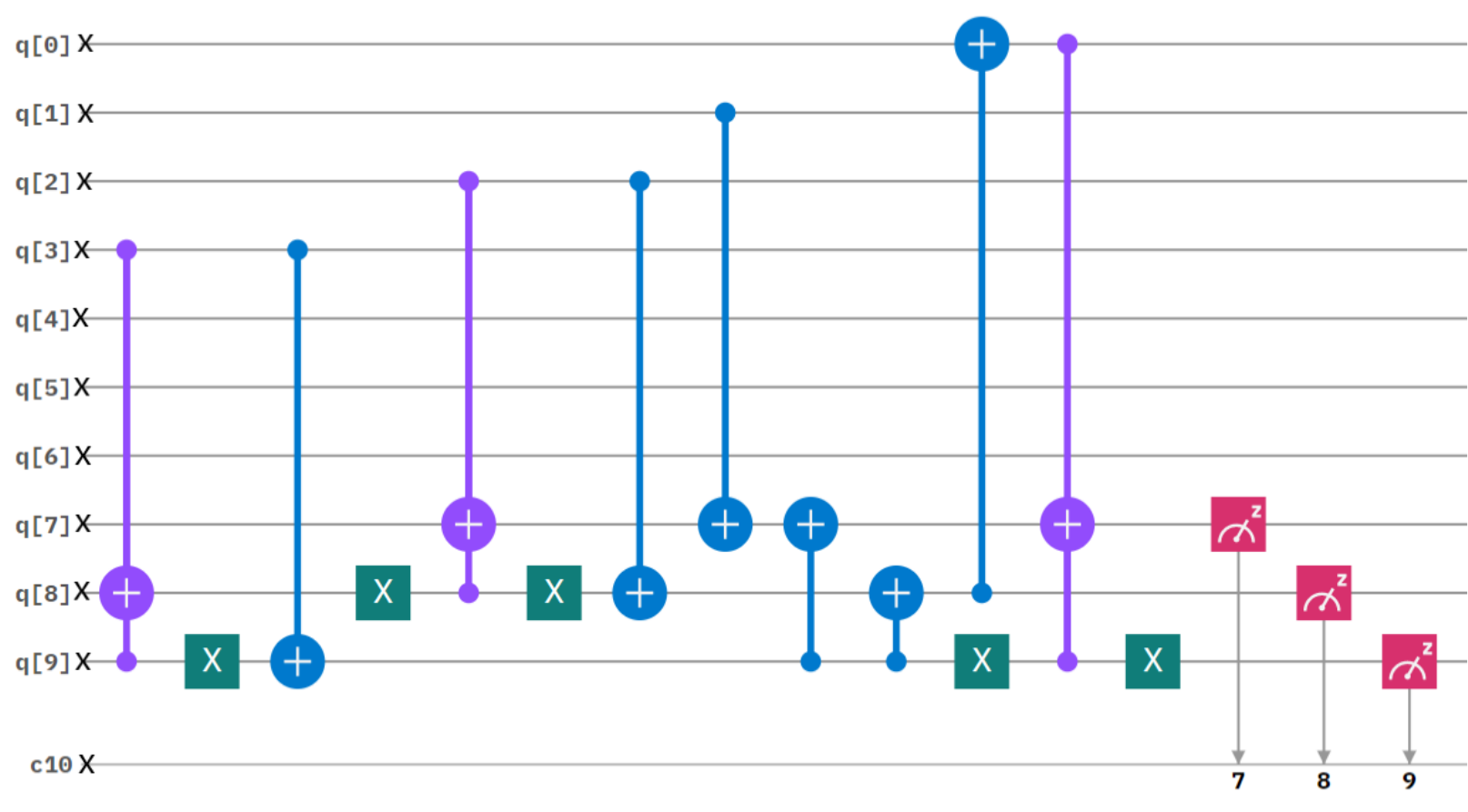}
    \caption{\textbf{Experimental circuit for teleportation of GHZ state.} \emph{The qubits q[1], q[2] and q[3] represent the position space. The qubits q[4], q[5] and q[6] represent Alice's coin and q[7], q[8] and q[9] represent Bob's coin. q[0] is an ancillary qubit.} }
    \end{subfigure}
     \label{qtc_Fig8}
\end{figure}

\begin{figure}[]
    \centering
    \includegraphics[scale=0.52]{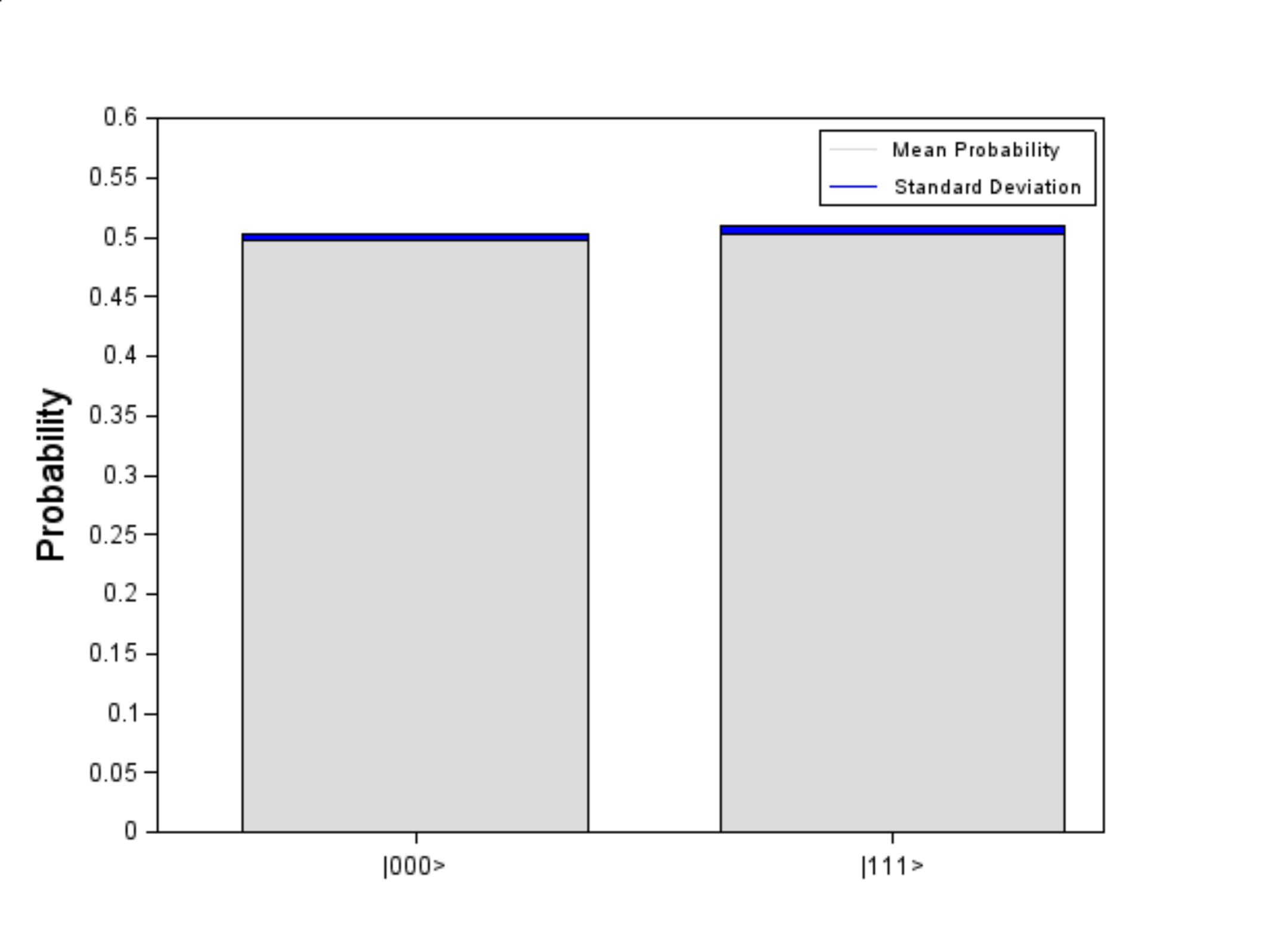}
    \caption{\textbf{Histogram.} \emph{The histogram shows the mean probabilities for $\ket{000}$ and $\ket{111}$ states and their standard deviations.}}
     \label{qtc_Fig9}
\end{figure}

The above circuit teleports a GHZ State as an example of a three-qubit state. The GHZ state is defined as, $\ket{GHZ}=\frac{\ket{000}+\ket{111}}{\sqrt{2}}$. The circuit is implemented on a 32-qubit simulator provided by IBM. For each simulation, 8192 shots have been used. Fig. \ref{qtc_Fig9} shows the graphical representation of data taken for 10 simulations. The measurements are in $Z$-basis.

Another interesting example of a three-qubit state is the W state. It has the following form, $$\ket{W}= \frac{\ket{001}+\ket{010}+\ket{001}}{\sqrt{3}}$$ It can be created using the following circuit \cite{qtc_raia}.

\begin{figure}[]
    \centering
    \includegraphics[scale=0.3]{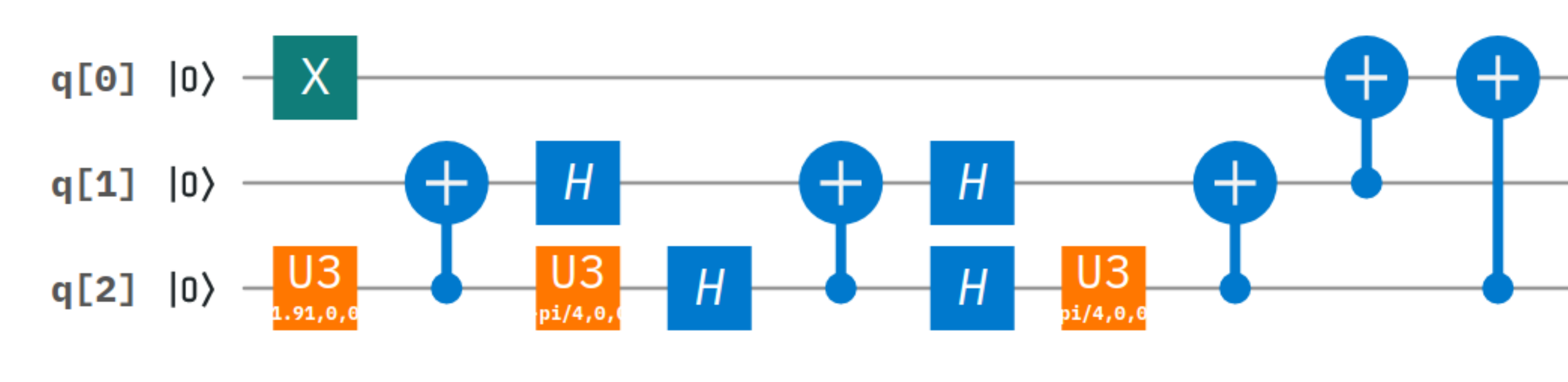}
    \caption{\textbf{Circuit for the creation of W state.}}
     \label{qtc_Fig10}
\end{figure}

\begin{figure}[H]
    \centering
    \includegraphics[scale=0.5]{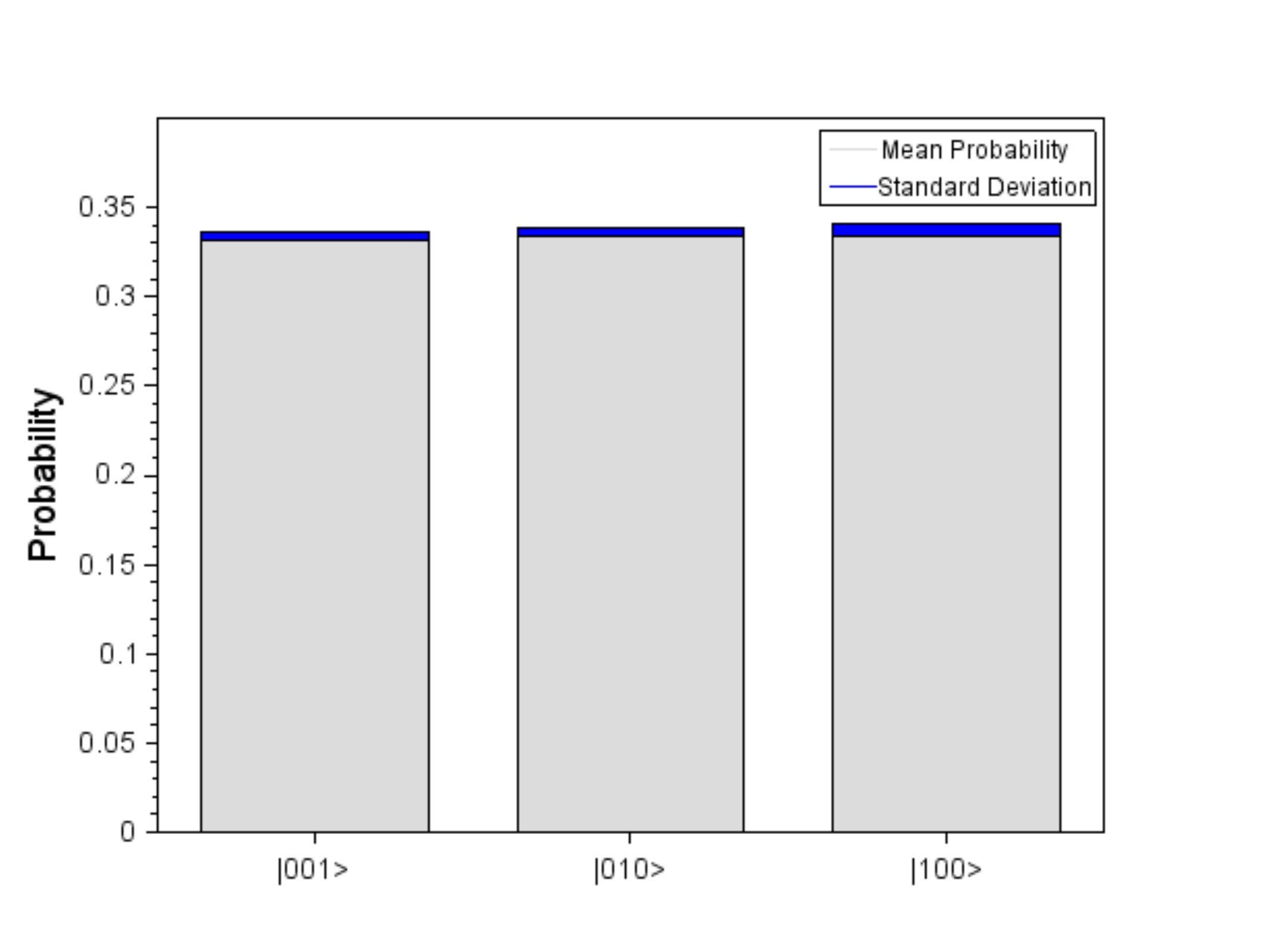}
   \caption{\textbf{Histogram.} \emph{This histogram shows the mean probabilities for $\ket{001}$, $\ket{010}$ and $\ket{100}$ states and their standard deviations.}}
  \label{qtc_Fig11}
\end{figure}

We have also teleported the W state using the circuits in Fig. \ref{qtc_Fig8} and Fig. \ref{qtc_Fig10}. Fig. \ref{qtc_Fig11} shows the graphical representation of data taken for 10 simulations. For each simulation, 8192 shots have been used. The measurements are in $Z$-basis.
 
\section{Conclusion \label{qtc_Sec7}}
We have demonstrated here an experimental procedure of quantum teleportation using coined quantum walks. Quantum state tomography has been performed to test the accuracy of our circuit for teleportation of a single-qubit state. It is observed that the teleportation is carried out with high fidelity. We have shown the teleportation of two-qubit and three-qubit states as well. In section \ref{qtc_Sec5.1}, we have provided a complete mathematical description of the teleportation scheme. It is important to note that the calculations shown in that section can be extended to perform the teleportation of any n-qubit system on a quantum computer provided the hardware is adequate. 

\section*{Acknowledgments}
\label{acknowledgments}
Y.C. and V.D. acknowledge the hospitality provided by IISER Kolkata. B.K.B. is financially supported by DST Inspire Fellowship. We thank Wang, Shang and Xue for their original contribution to the concept of using quantum walks for teleportation. We are extremely grateful to IBM and IBM QE project. The discussions and opinions developed in this paper are only those of the authors and do not reflect the opinions of IBM or IBM QE team.

\end{document}